# Cryogenic Ferroelectric Behavior of Wurtzite Ferroelectrics

Ruiqing Wang, Jiuren Zhou, *Member, IEEE,* Siying Zheng, Feng Zhu, Wenxin Sun, Haiwen Xu, Bochang Li, Yan Liu, Yue Hao, *Senior Member, IEEE,* and Genquan Han, *Senior Member, IEEE*

*Abstract*—This study presents the first experimental exploration into cryogenic ferroelectric behavior in wurtzite ferroelectrics. A breakdown field ($E_{BD}$) to coercive field ($E_C$) ratio of 1.8 is achieved even at 4 K, marking the lowest ferroelectric switching temperature reported for wurtzite ferroelectrics. Additionally, a significant evolution in fatigue behavior is captured, transitioning from hard breakdown to ferroelectricity loss at cryogenic temperatures. These findings unlock the feasibility for wurtzite ferroelectrics to advance wide temperature non-volatile memory.

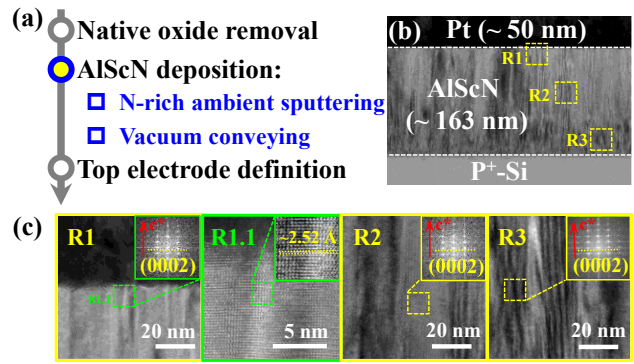

**Fig. 1.** (a) Key process flow for fabricating AlScN capacitors. (b-c) HRTEM images, demonstrating the obtained superior crystalline at both top interface and bulk regions.

## I. Introduction

The advancement of non-volatile memory (NVM) technologies capable of reliable operation across a broad temperature spectrum, spanning from cryogenic temperatures as low as 4 K to ultra-high temperatures exceeding 1200 K, remains a formidable challenge [1]-[3]. This constraint significantly impedes progress in critical domains such as aerospace systems, deep-space exploration, nuclear fusion, plasma technologies, and high-power laser systems [4], [5].

Wurtzite ferroelectrics, distinguished by their exceptionally high curie temperatures of approximately 1373 K, emerge as promising candidates for enabling NVM devices operable across these extreme thermal environments [6]. While their stability at elevated temperatures has been well-documented, showing ferroelectricity up to 873 K, their behavior under cryogenic conditions remains uncharted, associated with a questionable feasibility [3], [6], [7]. This knowledge gap stems primarily from material limitations at low temperatures, where the coercive field ($E_C$) approaches or exceeds the breakdown field ($E_{BD}$), thus usually leading to the loss of ferroelectricity [1]. To expand the low-temperature NVM applications of promising wurtzite materials, there is an urgent need to study the cryogenic ferroelectric behavior.

In this study, we bridge this gap by experimentally fabricating well-crystalline wurtzite AlScN and clarifying its feasibility for ferroelectric operating at cryogenic temperatures, down to 4 K. Beyond merely evaluating their ferroelectric polarization switching at cryogenic levels, the study also delves into critical reliability concerns in such a regime, particularly hard breakdown and fatigue, to provide a comprehensive understanding of their operational stability.

## II. Experiments

Fig. 1(a) outlines the fabrication process for wurtzite AlScN capacitors. The native oxide on a heavily doped P-type Si (001) wafer was removed by dilute hydrofluoric acid (DHF). An AlScN film was then sputtered at 200 °C using a single $Al_{0.8}Sc_{0.2}$ target, with high $N_2$ (160 sccm) and low Ar (32 sccm) flows. The vacuum conveying ensured the unoxidized AlScN film. Finally, the top electrode was defined with an area of 5024 μm². To prevent material failure at cryogenic temperatures, N-rich and oxygen-free processing conditions were used [7].

Fig. 1(b) illustrates the device structure, comprising a 163-nm-thick AlScN layer sandwiched between a Pt top electrode and a heavily doped P-type silicon bottom electrode. High-resolution transmission electron microscopy (HRTEM) images in Fig. 1(c) underscore the crystalline integrity of the

Manuscript received XXXX; revised XXXXX; accepted XXXXX. Date of publication XXXXXX; date of current version XXXXXXX. The authors acknowledge support from the National Science and Technology Major Project (No. 2022ZD0119002), the National Natural Science Foundation of China (Grant No. 92264101, 92464205, 62025402, 62090033), the Major Program of Zhejiang Natural Science Foundation (Grant No. LD25F040004), and the Postdoctoral Fellowship Program of China Postdoctoral Science Foundation (No. GZC20241309). The review of this paper was arranged by Editor XXXXX. (Corresponding author: Jiuren Zhou, Siying Zheng, and Feng Zhu)

R. Wang, J. Zhou, S. Zheng, W. Sun, H. Xu, B. Li, Y. Liu, Y. Hao, and G. Han are with School of Microelectronics, Xidian University, Xi'an, 710126, China, and also with Hangzhou Institute of Technology, Xidian University, Hangzhou, 311200, China. (e-mail: zhoujiuren@163.com; siying_zheng@163.com).

F. Zhu, with TRACE EM Unit and Department of Materials Science and Engineering, City University of Hong Kong, Hong Kong, 000000, China (e-mail: fengzhu@cityu.edu.hk).

Color versions of one or more figures in this letter are available at xx.

Digital Object Identifier xx

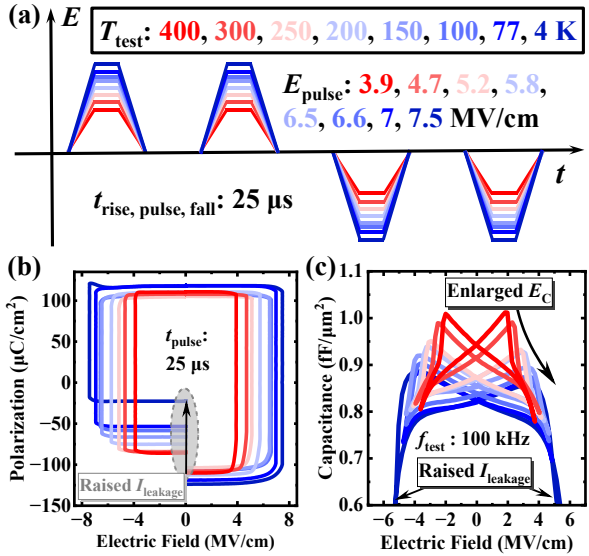

**Fig. 2** (a) PUND pulse scheme used for testing temperatures ($T_{test}$) ranging from 400 to 4 K. (b) PUND and (c) C-V measurements.

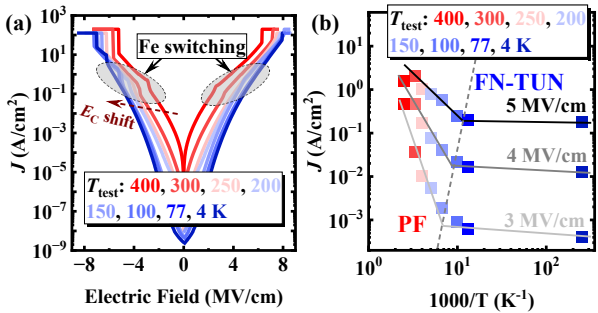

**Fig. 3** Breakdown properties of AlScN film, involving (a) TZDB test, (b) the leakage current as a function of 1000/T at different electric fields.

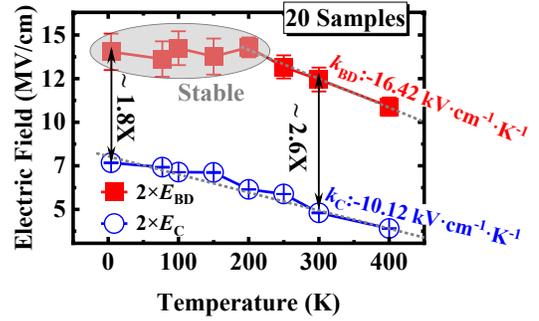

**Fig. 4.** Temperature dependence of the $E_{BD}$ and $E_C$ for 20 samples, with fitting $k_{BD}$ and $k_C$ constants and $E_{BD}/E_C$ of 1.8 at 4 K.

**Table I Temperature coefficient of $E_C$ in Ferroelectric Wurtzites**

|  | [15] | This work |
|---|---|---|
| Temperature range | 300 ~673 K | 400 ~ 4 K (cryogenic) |
| Temperature coefficient of ($E_C$) | - 4.50 kV•cm$^{-1}$•K$^{-1}$ | - 5.06 kV•cm$^{-1}$•K$^{-1}$ |

wurtzite ferroelectric layer. At the top interface (Region 1, R1), the samples exhibit intact crystalline with preserved direction and lattice structure, avoiding the typical oxidized interfacial layers seen in similar systems [8]. Diffraction analyses from bulk regions (Regions 2 and 3, R2 & R3) reveal almost uniform crystalline throughout the film, even at grain boundaries where nitrogen vacancies are prone to accumulate. These optimized fabrication measures effectively suppress oxidation at the interface and nitrogen vacancy formation within the bulk, ensuring superior crystalline and enhanced breakdown characteristics [9], [10].

## III. RESULTS AND DISCUSSION

The cryogenic ferroelectric polarization switching characteristics of wurtzite AlScN are detailed in Fig. 2. Fig. 2(a) illustrates the applied positive-up-negative-down (PUND) pulse trains, with a fixed pulse width of 25 μs, while the pulse amplitudes vary between 3.9 and 7.5 MV/cm, which accommodates the increased coercive field ($E_C$) induced by the temperature decrease, owing to the raised ferroelectric polarization switching barrier [11]. This measurement scheme thus ensured a consistent remnant polarization of our samples, approximately 100 μC/cm$^2$, across the whole tested temperature range, spanning from 400 to 4 K.

Fig. 2(b) presents the extracted polarization (P) versus (V) curves through the dynamic PUND test, demonstrating successful ferroelectric polarization switching at extreme-low temperature of 4 K, albeit with a significantly large $E_C$. The increase in $E_C$ of wurtzite ferroelectrics leads to an inevitable rise in the required external electric field, which in turn leads to an increase in leakage current, resulting in a widened polarization gap at an electric field near 0 MV/cm [11], [12]. Furthermore, Fig. 2(c) shows the quasi-static C-V curves for the same sample. The characteristic butterfly-shaped profile, along with the pronounced tailing effect, further confirms successful ferroelectric polarization switching at 4 K, as well as the obvious increases in $E_C$ and leakage current. Additionally, wurtzite ferroelectrics also exhibit a decrease in permittivity as the temperature lowers, which can be attributed to the suppression of dipoles orientation [13].

Focusing on the cryogenic operational reliability of wurtzite ferroelectrics, time-zero-dependent breakdown (TZDB) measurements were conducted. Fig. 3(a) presents the corresponding I-V curves at various temperatures. The small humps and abrupt jumps in the I-V curves mean the ferroelectric polarization switching ($E_C$ points) and hard breakdown (breakdown electric field points, $E_{BD}$ points) of AlScN capacitors, respectively. Although the $E_C$ of wurtzite ferroelectrics increases continuously with decreasing temperature, it remains smaller than the $E_{BD}$, which guarantees the reliability of ferroelectric polarization switching in cryogenic temperatures. Notably, a convergence between $E_{BD}$ and $E_C$ is observed as the temperature decreases [14], [15].

To investigate the underlying conduction mechanisms with temperature transformation, Fig. 3(b) plots the leakage current of the samples as a function of 1000/T. At higher temperatures, the leakage current exhibits a strong temperature dependence; however, this dependence weakens at lower temperatures, indicating a transition in the conduction mechanism from Poole-Frenkel (PF) hopping to Fowler-Nordheim (FN) tunneling [16]. Especially, such a low knee point observed around 150-200 K in the current wurtzite ferroelectrics, is much smaller than the counterparts of fluorite ferroelectrics, around 400 K [16], underscoring the necessity of wurtzite ferroelectrics for further enhancement in crystalline, towards the advanced applications at even lower temperatures.

Shifting to the temperature-dependent characteristics of the

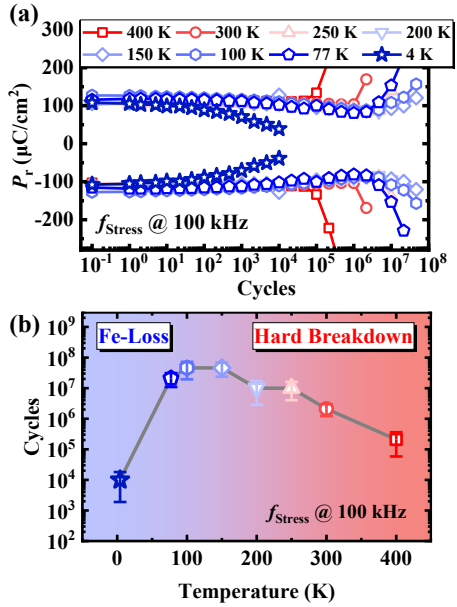

**Fig. 5** (a) Endurance characteristics of the ferroelectric AlScN capacitor under wide-ranging temperatures. (b) Cumulative analyses of endurance, showing a distinct transition with temperature change.

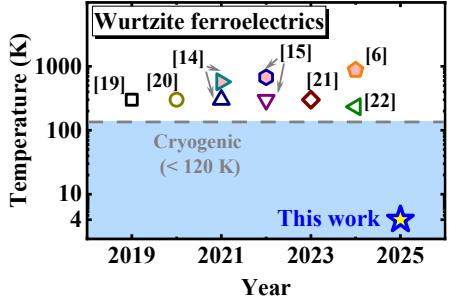

**Fig. 6.** Comparison of extreme temperature characteristics of wurtzite ferroelectrics, emphasizing the record-low temperature of 4 K achieved in this work.

[18], requiring further investigation into nitrogen vacancy optimization and the underlying mechanisms governing fatigue properties.

Fig. 6 benchmarks the extreme research temperatures for the reported wurtzite ferroelectrics so far, highlighting our pioneering entry into the cryogenic temperature regime, with reliable ferroelectric polarization switching down to 4 K [6], [14], [15], [19]-[22]. This milestone underscores the significant potential of wurtzite ferroelectrics for the advance wide-temperature applications.

## IV. CONCLUSION

For the first time, the operational temperature of wurtzite ferroelectrics has been extended to the cryogenic regime, remarkably down to 4 K. Coupled with comprehensive reliability investigations, this work paves the way for significant advancements in non-volatile memory technologies tailored for extreme wide-temperature environments.

$E_{BD}$ and $E_C$, Fig. 4 presents statistical data on $E_C$ and $E_{BD}$ as a function of temperature, which are extracted from quasi-static $C$-$V$ and $I$-$V$ curves. The device counts are 20. The $E_C$ exhibits a monotonic increase as the temperature decreases from 400 to 4 K, with the cryogenic temperature coefficient of $E_C$ reported for the first time, as - 5.06 kV·cm$^{-1}$·K$^{-1}$ (half of $k_C$ in Fig. 4), as summarized in the following Table I. In contrast, $E_{BD}$ displays a distinct trend: it initially increases with decreasing temperature before stabilizing, with a knee point around 200 K, which corresponds well to the conduction mechanism transition illustrated in Fig. 3(b). Remarkably, even at 4 K, our samples maintain a robust $E_{BD}/E_C$ ratio of 1.8, enabling reliable ferroelectric polarization switching in wurtzite ferroelectrics.

Furthermore, the cryogenic fatigue characteristics of wurtzite ferroelectrics are examined in Fig. 5, with a stress frequency of 100 kHz and varying pulse amplitudes applied depending on the test temperatures. The competitive endurance characteristics, up to $2\times10^6$ @ 300 K, is obtained in our samples. More importantly, Fig. 5 reveals a distinct transition in fatigue behavior: breakdown dominates at high temperatures, whereas ferroelectricity loss becomes prominent at cryogenic temperatures, which both strongly relate to nitrogen vacancy motion and ferroelectric domain pinning [17],